\documentclass[12pt]{article}

\usepackage{color}
\usepackage{amsfonts}
\usepackage{amsmath}  
\usepackage{graphics}
\usepackage{multirow}
\usepackage{threeparttable}
\usepackage{verbatim}
\usepackage{latexsym}
\usepackage{booktabs}
\usepackage{algorithm}
\usepackage{algorithmic}
\usepackage{amsmath}
\usepackage{txfonts}
\usepackage{amssymb}
\usepackage{lscape}
\usepackage{subfigure}
\usepackage{multicol}
\usepackage{float}
\usepackage{float}
\usepackage[caption = false]{subfig}
\usepackage[demo]{graphicx}
\usepackage{chngcntr}
\usepackage{psfrag}
\usepackage{amssymb} 
\usepackage[hmargin = 1.2in, vmargin = 1.4in]{geometry}

\title{Forecasting multivariate longitudinal binary data with marginal and marginally specified models}
\author{\"{O}zg\"{u}r Asar$^{1}$\footnote{Correspondence to: \"{O}zg\"{u}r Asar, Lancaster Medical School, Faculty of Health and Medicine, Lancaster
\hspace*{1.6em} University, LA1 4YG, UK. E-mail: o.asar@lancaster.ac.uk} \ \ and \"{O}zlem \.{I}l$\mbox{k}^{2}$ \\ \\
$^1$ CHICAS, Lancaster Medical School, Lancaster University, UK \\
$^2$ Department of Statistics, Middle East Technical University, Turkey}

\begin{document}

\maketitle

\begin{abstract}

Forecasting with longitudinal data has been rarely studied. Most of the available studies are for continuous response and all of them are for univariate response. In this study, we consider forecasting multivariate longitudinal binary data. Five different models including simple ones, univariate and multivariate marginal models, and complex ones, marginally specified models, are studied to forecast such data. Model forecasting abilities are illustrated via a real life data set and a simulation study. The simulation study includes a model independent data generation to provide a fair environment for model competitions. Independent variables are forecast as well as the dependent ones to mimic the real life cases best. Several accuracy measures are considered to compare model forecasting abilities. Results show that complex models yield better forecasts.

\end{abstract}


\section{Introduction}

Longitudinal data comprise measurements which are taken repeatedly over time from same individuals/firms/units/cases/animals. This type of data is common in many research areas, e.g. medical studies, clinical trials, economical studies, social sciences, psychiatry, educational and behavioral sciences, industry, etc. Longitudinal data have many advantages compared to both time series and cross sectional data (Diggle et al., 2002; Ilk, 2008).

Observations in longitudinal data are typically dependent. In related studies, often multiple response variables of each study subject are collected, which yields multivariate longitudinal data. Multivariate longitudinal data consist of three dependence structures: within, between and cross response dependencies and these structures should be taken into account to have valid statistical inferences. 

As an example for multivariate longitudinal data, we will consider the Mother's Stress and Children's Morbidity (MSCM) study data set (Alexander and Markowitz, 1986). In this study, 167 mothers and their pre-school children were followed through 28 days. At each day, mothers' stress (0=absence, 1=presence) and their children's illness (0=absence, 1=presence) statuses were recorded, yielding a bivariate longitudinal binary data. In addition to these response variables, some demographic and family information, e.g. employment status of mothers (0=unemployed, 1=employed), health status of children at baseline (0=very poor/poor, 1=fair, 2=good, 3=very good) and the size of their household (0=2-3 people, 1=more than 3 people) were collected. 

There are three traditional models for longitudinal data analysis: marginal, transition and random effects models (Diggle et al., 2002). Recently, marginally specified (marginalised) models (Heagerty, 1999, 2002) have become popular due to their advantages over the traditional ones. For instance, they incorporate marginal, transition and random effects inferences at the same time (Ilk and Daniels, 2007; Asar et al., 2014). Additionally, they secure the robustness of the marginal mean parameters under dependence structure misspecification (Heagerty and Kurland, 2001). Nonetheless, they require more computational time to obtain the parameter estimates compared to the aforementioned traditional ones.

Forecasting might be regarded as the prediction of future events. It might be a life-saver and/or increase quality of life, e.g. in medical and social studies. For instance, we can forecast the mothers' stress and their children's illness statuses in the MSCM study. Moreover, we can incorporate the relationships of the response variables with the explanatory ones while achieving these forecasts via the longitudinal models. These would help taking precautions based on subject and/or sub-group characteristics (e.g. employed vs. unemployed mothers). Eventually, infants might be precluded to have illness and mothers' life quality could be increased.

In forecasting, one of the key suggestions is constructing simple models. The other suggestion is forecasting data for near future, as the forecast accuracy might decrease when forecasting data for far future. For instance, Diggle (1990, p. 189) showed that the variance of the forecast increases as the lag between forecast and available data time points increases. 

Forecasting is common in time series literature, e.g. see Nobre et al. (2001) and Burkom et al. (2007), but it is rare for longitudinal data (Baltagi, 2008). This might be due to the spans of these two research areas; it is well-known that the first has a longer history. Moreover, in some time series studies, e.g. in applied macroeconomics and financial econometrics, the main aim of data collection and model building is forecasting Harris and Sollis (2003, p. 10). On the other hand, in longitudinal studies, the main interest is mostly on understanding the relationships between the dependent and the independent variables and/or drawing subject-specific inferences, by measuring the change. Nevertheless, forecasting with longitudinal data might be more informative and richer compared to the one with time series data. For example, longitudinal data models, specifically random effects models, allow subject-specific forecasts in addition to the explanation of how the forecasts are related to independent variables.

Most of the available forecasting studies for longitudinal data have an econometric perspective; a literature survey of such studies up to 2008 could be found in Baltagi (2008). Moreover, most of these studies are for continuous response and all of them are for univariate response. For instance, Baadsgaard et al. (2004) considered forecasting health statuses of pig herds. Related data consisted of 15 Danish pig herds with 12 month follow-ups. They mainly considered the comparison of the forecasting abilities of single moving average method and a Bayesian state space model. Related results showed that these methods performed similar. Frees and Miller (2004) concentrated on forecasting Wisconsin Lottery Sales. Data were available for 40 weeks (April, 1998 - January, 1999) for 50 different postal codes located in Wisconsin. They mainly considered simple and complex versions of random effects models. Results indicated that complex models did not outperform the simpler ones in terms of forecasting, even though they did so in terms of model building. Aslan (2010) considered a simulation study on forecasting univariate longitudinal binary data. The author mainly concentrated on the comparison of forecasting abilities of 21 different forecasting methods including simple methods, e.g. moving and non-moving mean, median and mode and complex models, e.g. marginal, transition, random effects and marginalised transition models (Heagerty, 2002). This study considered a model independent data simulation scenario which permits the models to fairly compete. It further considered forecasting the covariates. Results showed that random intercept and transition models performed the best in terms of forecasting. 

In this study, we considered forecasting \emph{multivariate longitudinal binary data}. We mainly concentrated on comparing forecasting performances of univariate and multivariate marginal models and two marginally specified models. Our main motivating question of interest was whether marginalised models yield better forecasts compared to the simpler ones. We believe that answering this question is important since marginalised models require more time and effort during both model building and forecasting, e.g. they require forecasting the time varying parameters and/or the use of iterative numerical methods. MSCM data set were used to illustrate this comparison in real life. Moreover, results were compared via a simulation study. Following Aslan (2010), we considered a model independent data generation process and we forecast the independent variables as well.

The remainder of the paper is organised as follows. In Section 2, we provide the details of the models, related forecasting methodologies and the accuracy measures. Section 3 provides the details of the MSCM data set and the related forecasting results. Section 4 provides the details of data generation and the forecasting results of the simulation study. The paper is closed with conclusion and discussion.\\

\section{Methods}

\subsection{Models}

We mainly considered five different models to forecast multivariate longitudinal binary data. Here, we briefly presented the modelling frames and some of their distinguishing features due to page limits. Details can be found in the related references cited below. \\

\subsubsection{Univariate marginal models}

The modelling framework of univariate marginal models (UMM) is given by

\begin{equation}
\label{eq:umm}
\mbox{logit} \ \mbox{P} (Y_{it}=1|X_{it})=X_{it} \ \beta ,
\end{equation}

\noindent where $Y_{it}$ is the (univariate) response for subject $i$ $(i=1, \ldots, N)$ at time $t$ $(t=1, \ldots, T)$; $X_{it}$ are the associated set of covariates, $\beta$ are the regression parameters to be estimated and $\mbox{logit}$ is the log of odds. A popular approach to obtain the estimates of $\beta$, $\hat{\beta}$, is the generalized estimating equations (GEE; Diggle et al., 2002; Liang and Zeger, 1986; Zeger and Liang, 1986). In this approach, a working variance-covariance structure is used for the repeated observations to take the within response dependency into account during the parameter estimation process. 

UMM considers building separate univariate models for each longitudinal responses one by one, with possibly different set of covariates for different responses. In other words, it only considers within response dependency and ignores the between and cross response dependencies. 

\subsubsection{Multivariate marginal models with response specific parameters}

Multivariate marginal models with response specific parameters (MMM1; Shelton et al., 2004; Asar and Ilk, 2013) is an extension of UMM to multivariate response data. The related model formulation is given by

\begin{equation}
\label{eq:mmm1}
\mbox{logit} \ \mbox{P} (Y_{itj}=1|X_{it})=X_{it} \ \beta_j,
\end{equation}

\noindent where $Y_{itj}$ is the $j$th $(j=1, \ldots, k)$ response for subject $i$ at time $t$, $X_{it}$ are the common set of covariates for multiple responses and $\beta_j$ are the response specific regression parameters. 

MMM1 uses GEE for parameter estimation and takes within, between and cross response dependencies into account via working variance-covariance structure of the multiple responses. \\

\subsubsection{Multivariate marginal models with shared regression parameters}

Asar and Ilk (2014) proposed multivariate marginal models with shared regression parameters (MMM2) by extending the MMM1 formulation in terms of covariate set specification and regression parameter assumption. The modelling framework of MMM2 is given by

\begin{equation}
\label{eq:mmm2}
\mbox{logit} \ \mbox{P} (Y_{itj}=1|X_{itj})=X_{itj} \ \beta ,
\end{equation}

\noindent where $X_{itj}$ are the response specific set of covariates and $\beta$ are the regression parameters that are shared across multiple responses. We can still allow multiple responses to have their own intercepts by including response type indicator variable(s) in the design matrix. Similarly, we can allow them to have their own slopes by including the interactions of these indicator variables with covariates. By this setup, we can build more parsimonious multivariate marginal models compared to MMM1 as well as equivalent ones. Similar to MMM1, MMM2 uses GEE for parameter estimation and considers the aforementioned three dependence structures via working variance-covariance structure of the multivariate responses. \\

\subsubsection{Marginalised multivariate random effects models}

Marginalised multivariate random effects models (MMREM) were proposed by Lee et al. (2009) to analyse multivariate longitudinal binary data. The framework includes two level logistic regression models which are given by

\begin{eqnarray}
\label{eq:mmremlev1}
&& \mbox{logit} \ \mbox{P} (Y_{itj}=1|X_{it})=X_{it} \ \beta_j , \\
\label{eq:mmremlev2}
&& \mbox{logit} \ \mbox{P} (Y_{itj}=1|X_{it},b_{itj})=\Delta_{itj}+b_{itj}.
\end{eqnarray}

\noindent Here, $b_{itj}$'s are the subject, time and response specific random effects. The random effects of subject $i$ are assumed to follow a multivariate normal distribution, i.e., $b_i=(b_{i11}, \ldots, b_{i1k}, \ldots,$ $ b_{iT1}, \ldots, b_{iTk})^{T}$ $\sim N(0, \Sigma)$ with $\Sigma=\Sigma_1 \otimes \Sigma_2$, where $\otimes$ corresponds to Kronecker product. On the one hand, $\Sigma_1$ is a within response correlation matrix having an AR-1 structure, i.e., structured by only one transition parameter, $\gamma$. On the other hand, $\Sigma_2$ is a variance-covariance matrix of multiple responses structured by $(k \times (k+1))/2$ covariance parameters. $\Delta_{itj}$ is the subject, time and response specific intercept that takes the non-linear relationship between \eqref{eq:mmremlev1} and \eqref{eq:mmremlev2} into account. It is deterministic function of other model parameters and obtained by solving the following convolution equation: 

\begin{equation}
\label{eq:mmremconv}
\mbox{P} (Y_{itj}=1|X_{it})= \int_{b_{itj}} \mbox{P} (Y_{itj}=1|X_{it},b_{itj}) \ d F(b_{itj}).
\end{equation}  

The orthogonalization of the random effects by setting $b_i=\Sigma_1^{1/2} \otimes \Sigma_2^{1/2} C_i$, where $C_i $ is a $(T \times k) \times 1$ matrix with identical elements of $z_i$, where $z_i \sim N(0,1)$, yields \eqref{eq:mmremlev2} to have the following re-parametrised form:

\begin{equation}
\label{eq:mmremnewlev2}
\mbox{logit} \ \mbox{P} (Y_{itj}=1|z_i,X_{it})=\Delta_{itj}+r^{k(t-1)+j}C_i ,
\end{equation}

\noindent where $r^{k(t-1)+j}$ is the $(k(t-1)+j)$th row of $\Sigma^{1/2}=\Sigma_1^{1/2} \otimes \Sigma_2^{1/2}$.

MMREM takes the aforementioned three dependence structures into account. The parameter estimates are obtained by maximum likelihood estimation (MLE). Empirical Bayesian estimators of $z_i$ can be found in Asar (2012, Chapter 2.4.3) which were not originally derived in Lee et al. (2009). \\

\subsubsection{Probit normal marginalised transition random effects models}

Asar et al. (2014) proposed probit normal marginalised transition random effects models (PNMTREM) by extending the work of Ilk and Daniels (2007). Here, we only considered first order PNMTREM, i.e., PNMTREM(1), during our forecasting studies. General model specifications could be found in these references. The modelling formulation of PNMTREM(1) for $t \geq 2$ is given by

\begin{eqnarray}
\label{eq:pnmtremlev1}
&& \mbox{P} (Y_{itj}=1|X_{itj})=\Phi( X_{itj} \beta ) , \\
\label{eq:pnmtremlev2}
&& \mbox{P} (Y_{itj}=1|y_{i,t-1,j},X_{itj})=\Phi(\Delta_{itj}+ \alpha_{t} Z_{itj} y_{i,t-1,j}) , \\
\label{eq:pnmtremlev3}
&& \mbox{P} (Y_{itj}=1|y_{i,t-1,j},X_{itj},b_{it})=\Phi(\Delta_{itj}^*+\lambda_j  b_{it}) . 
\end{eqnarray}

\noindent Here, $\alpha_{t}$ is a transition parameter vector that captures the relationship between $Y_{itj}$ and $Y_{i,t-1,j}$. $Z_{itj}$ is typically a subset of covariates, $X_{itj}$, $b_{it}$'s are the subject and time specific random effects with $b_{it}\sim N(0,\sigma_t^2)$ and $b_{it}=\sigma_t z_i$, $\lambda_j$'s are the response specific parameters with $\lambda_1=1$ for identifiability reasons and $\Phi(.)$ is the cumulative distribution function of standard normal. $\Delta_{itj}$ is the subject, time and response specific intercept that takes the non-linear relationship between \eqref{eq:pnmtremlev1} and \eqref{eq:pnmtremlev2}. It is a deterministic function of model parameters and can be obtained by solving

\begin{equation}
\label{eq:pnmtremmargcons}
\mbox{P} (Y_{itj}=1|X_{itj})=\sum\limits_{y_{i,t-1,j}} \mbox{P} (Y_{itj}=1|y_{i,t-1,j},X_{itj}) \mbox{P} (Y_{i,t-1,j}|X_{i,t-1,j}).
\end{equation}

\noindent Similarly, $\Delta_{itj}^*$ takes the non-linear relationship between \eqref{eq:pnmtremlev2} and \eqref{eq:pnmtremlev3} and can be obtained by solving

\begin{equation}
\label{eq:pnmtremconv}
\mbox{P} (Y_{itj}=1|y_{i,t-1,j},X_{itj})= \int_{b_{it}} \mbox{P} (Y_{itj}=1|y_{i,t-1,j},X_{itj},b_{it})dF(b_{it}).
\end{equation}

Since no history data are available at hand, a different model is assumed for baseline time point $(t=1)$. It also has a marginalised modelling structure and related framework is given by

\begin{eqnarray}
\label{eq:pnmtrembase1}
&& \mbox{P} (Y_{i1j}=1|X_{i1j})=\Phi(X_{i1j} \beta^*),\\
\label{eq:pnmtrembase2}
&& \mbox{P} (Y_{i1j}=1|X_{i1j},b_{i1})=\Phi(\Delta_{i1j}^*+\lambda_j^* b_{i1}),
\end{eqnarray}   

\noindent where $\beta^*$ captures the relationship between the covariates and the mean response at $t=1$, $\Delta_{i1j}^*$ takes the non-linear relationship between \eqref{eq:pnmtrembase1} and \eqref{eq:pnmtrembase2}, to be calculated by solving a convolution equation which can be seen by setting $t=1$ and omitting $Y_{i,t-1,j}$ in \eqref{eq:pnmtremconv}, $b_{i1}$ is the subject specific random effects with $b_{i1}\sim N(0,\sigma_1^2)$ and $b_{it}=\sigma_1 z_i$ and $\lambda^*_j$ is the response-specific parameters with $\lambda^*_1=1$ for identifiability.

Parameters of PNMTREM(1) are estimated by MLE and the modelling framework considers within and between response dependencies. However it does not directly take the cross response dependencies into account. The $z_i$'s are estimated by Empirical Bayesian estimation and the details can be found in Asar (2012, Appendix B.5). \\

\subsection{Forecasting methodologies}

\subsubsection{Independent variables}

Independent variables in longitudinal data might be time varying or time invariant. Time invariant variables are not needed to be forecast, since related observations are constant over time. On the other hand, future values of the former type are random variables indeed and need to be forecast. Nonetheless, forecasting is not needed for some time varying variables which are deterministic functions of time, e.g. age. In forecasting literature, people usually assume that the independent variables are known and only forecast the dependent ones. However, in real life the time variant covariates are unknown as well and should be forecast.

It might be beneficial to first consider forecasting the independent variables, since the forecasting methodologies of dependent variables rely on complete design matrices, i.e., $\hat{X}_{it}$ at $t=(T+1), \ldots, (T+m)$ assuming we intend to do $m$ step ahead forecasting.

Methods relying on the history of independent variables might be the best choices to forecast longitudinal independent variables. Alternative methods, e.g. the ones accommodating the relationship with other independent variables might be considered, but there are some difficulties while using such methods. For instance, if the related variables include time varying ones, we need their future values as well. These methods might not be the best choice anyway, since we expect low correlations between the independent variables. 

All the independent variables were time invariant in the MSCM data set. On the other hand, we assumed time varying independent variables in the simulation study. All of the independent variables were assumed to follow Gaussian distribution and the correlations among these variables were assumed to be low. We mainly considered first and second order transition models, TM(1) and TM(2), to forecast the independent variables in the simulation study. Other methods that rely on the history of the independent variables were considered in Aslan (2010), e.g. moving average and random effects models. But the author reported that these alternatives did not yield better forecasts compared to TM(1) and TM(2).

The modelling formulation of TM(1) for Gaussian data is given by

\begin{equation}
\label{eq:tm1}
X_{itl}=\beta_0+\beta_1 X_{i,t-1,l} + \epsilon_{itl},
\end{equation}

\noindent where $\epsilon_{itl} \sim N(0, \sigma^2)$ and $l$ is the covariate index for any time varying covariate. 

Similarly, the framework of TM(2) can be given by

\begin{equation}
\label{eq:tm2}
X_{itl}=\beta_0+ \beta_1 X_{i,t-1,l} + \beta_2 X_{i,t-2,l} + \epsilon_{itl}.
\end{equation}

The forecasting methodology of TM(1) can be briefly explained as obtaining $\hat{\beta}_0$ and $\hat{\beta_1}$, by using the available data $(t=1, \ldots, T)$ and replacing them in \eqref{eq:tm1} together with $\hat{X}_{i,t-1}$ for $(T+2), \ldots, (T+m)$. Forecasting methodologies of TM(2) is similar to the one for TM(1). \\

\subsubsection{Dependent variables}

\noindent The forecasting methodologies of UMM, MMM1 and MMM2 are similar to each other. Therefore, here we only illustrate the one for UMM. It can be summarised as follows: Obtain the estimates of $\beta$, $\hat{\beta}$, based on the available data for each response, possibly with different sets of independent variables. Then the forecasts of the success probabilities, $\hat{p}_{it}=\hat{\mbox{P}} (Y_{it}=1|\hat{X}_{it})$, can be obtained by replacing them in \eqref{eq:umm} together with $\hat{X}_{it}$. 

Forecasting with MMREM and PNMTREM(1) are more complex compared to the above models due to their complex structures. For instance, MMREM and PNMTREM(1) include time varying parameters, i.e. $\Delta_{itj}$ in MMREM and $\alpha_{t}, \sigma^2_t, \Delta_{itj}$ and $\Delta_{itj}^*$ in PNMTREM(1) and we need to forecast these parameters as well. Related forecasting methodologies are quite different and illustrated below separately.\\

\noindent - MMREM

\begin{enumerate}
\setlength{\itemsep}{-0.05in}
\item Obtain the estimates of $\beta_j$, $\Sigma_{1}^{1/2}$, $\Sigma_{2}^{1/2}$ and $z_i$. 
\item Extend $\Sigma_1^{1/2}$ from a $T \times T$ matrix to a $(T+m) \times (T+m)$ matrix, $\Sigma_{1,new}^{1/2}$, based on the estimate of the transition parameter, $\gamma$. Since $\Sigma_2^{1/2}$ is time invariant, no extension is needed for this matrix.
\item Obtain $\Sigma_{new}^{1/2}=\Sigma_{1,new}^{1/2} \otimes \Sigma_2^{1/2}$.  
\item Obtain the forecast of $\Delta_{itj}$, $\hat{\Delta}_{itj}$, by solving the non-linear equation given in \eqref{eq:mmremconv} via 40-points Gauss-Hermite Quadratures and Newton-Raphson algorithm in terms of $\Delta_{itj}$ based on $\hat{X}_{it}$, $\hat{\beta}_j$, $\hat{\Sigma}_{2}^{1/2}$ for $t=(T+1), \ldots, (T+m)$. Note that \eqref{eq:mmremconv} is free of $\gamma$ (Lee et al., 2009, p. 1287) and therefore, we do not need $\hat{\gamma}$ to obtain $\hat{\Delta}_{itj}$.  
\item Obtain the forecast of the success probability, $\hat{p}_{itj}=\hat{\mbox{P}} (Y_{itj}=1|\hat{z}_i,\hat{X}_{it})$ by using \eqref{eq:mmremnewlev2}.
\end{enumerate}

We considered different methodologies while calculating $\hat{p}_{itj}$. Whereas the first one (MMREM1) relied on using the columns $1$ to $(T+m)$ of $\Sigma_{1,new}^{1/2}$, the second (MMREM2) relied on using columns $(T+1)$ to $(T+m)$ of it. Alternative models were also considered by using different estimation methods for $z_i$'s. For instance, the third model (MMREM3) relied on generating $z_i$'s from independent standard normal distributions $K$ times, and calculating $K$ different $\hat{p}_{itj}$'s for each subjects and taking median of these quantities. We preferred median, since empirical investigations of success probabilities for randomly selected subjects suggested highly right-skewed distributions for the MSCM data set. We suggested studying percentage differences in the accuracy measures for successive $K$ values, e.g. $K=30$ vs. $K=50$, $K=50$ vs. $K=80$ and so on, and selecting the one for which the percentage changes were reasonable, i.e., causing little changes which were not worth increasing $K$. The last method with MMREM (MMREM4) relied on only using $\hat{\Delta}_{itj}$ while calculating $\hat{p}_{itj}$'s, i.e., taking $z_i=0$. This simplified the calculations at a cost of rather unrealistic assumption that all subjects are average.\\

\noindent - PNMTREM(1)

\begin{enumerate}
\setlength{\itemsep}{-0.05in}
\item Obtain the estimates of $\beta$, $\alpha_{t}$, $\sigma_t$, $z_i$.
\item Obtain forecasts of $\alpha_{t}$ and $\sigma_t$ for $t= (T+1), \ldots, (T+m)$ by exponential smoothing methods (Hyndman and Khandakar, 2008) for the MSCM data set (8 time points for model building) and simple moving averages method for the simulated datasets (4 time points for model building).
\item Obtain the forecast of $\Delta_{itj}^*$, $\hat{\Delta}_{itj}^*$, as given in \eqref{eq:pnmtremconv}.
\item Obtain the forecast of $\hat{p}_{itj}$ by using \eqref{eq:pnmtremlev3}.
\item Dichotomise $\hat{p}_{i,t-1,j}$ by considering a classification rule such that $\hat{Y}_{i,t-1,j}=1$ if $\hat{p}_{i,t-1,j} \geq c_j$ and 0 otherwise for $t=(T+2), \ldots, (T+m)$ where $c_j$ is a cutoff value.
\end{enumerate}

The dichotomization in step 5 was only necessary in PNMTREM(1), since we need history data in \eqref{eq:pnmtremlev2}. Eight different forecasting methodologies were considered for PNMTREM(1). These methodologies were produced by mainly combining estimation of $z_i$ while calculating the $\hat{p}_{itj}$ and handling $Y_{i,t-1,j}$ in \eqref{eq:pnmtremlev2} at $day=26, 27$ and $28$. Whereas the former included two options, using the Empirical Bayes estimates of $z_i$, $\hat{z}_i$, and assigning $z_i=0$, the latter included four options, setting $c_j=0.5$ to dichotomise $\hat{p}_{i,t-1,j}$, using the true values of $Y_{i,t-1,j}$ instead of dichotomizing $\hat{p}_{i,t-1,j}$, using empirical proportions of the responses as the $c_j$ and simulating the responses from independent Bernoulli distributions with success probabilities, $\hat{p}_{i,t-1,j}$. For instance, combining the two options of estimation of $z_i$ and the first option of handling $Y_{i,t-1,j}$ yielded the following two forecasting methodologies: using $\hat{z}_i$ and setting $c_j=0.5$ and assuming $z_i=0$ and setting $c_j=0.5$. We specifically call these methodologies as PNMTREM1 and PNMTREM2, respectively. Remaining six methodologies were produced by following similar strategies. We preferred not to give specific names to them to avoid unnecessary abbreviations, since we reported only the results of PNMTREM1 and PNMTREM2 in the third section. Nonetheless, all the results will be discussed in the same section. \\

\subsection{Accuracy measures}

We considered several accuracy measures to compare the model performances. Specifically, two different accuracy measures were considered for dependent binary variables, and two different accuracy measures were considered for independent continuous variables. Below we discuss these measures. \\

\subsubsection{Binary data accuracy measures}

\noindent The first accuracy measure we considered for binary data was the expected proportion of correct prediction (ePCP) which was proposed by Herron (1999). ePCP considers the calculation of average probability of estimating the actual observations, i.e., it considers $(1-\hat{p}_{itj})$ when $Y_{itj}=0$ and $\hat{p}_{itj}$ when $Y_{itj}=1$, where $\hat{p}_{itj}$ is the estimated success probability. The calculation of ePCP is given by

\begin{align} 
\mbox{ePCP}_{tj} &=\frac{1}{N} \sum_{i=1}^{N} \left( y_{itj} \ \hat{p}_{itj} + (1-y_{itj}) \ (1-\hat{p}_{itj}) \right).
\end{align}

\noindent The interval in which the possible values of ePCP lie is $[0,1]$ and larger values indicate better performance.

The second binary accuracy measure we considered was the area under the receiver operating characteristics (AUROC) curve. AUROC considers all the possible $c_j$ (between 0 and 1) and dichotomises the estimated success probabilities as 0 or 1 with respect to these values. Then, the area under the curve which is drawn by placing false positive rate (FPR) on the x-axis and true positive rate (TPR) on the y-axis is considered as the corresponding accuracy measure. Here, while TPR is calculated by the ratio of the number of cases which is assigned as positive (1, here) and is actually observed to be 1 to the total number of actual positives, FPR is calculated by the ratio of the number of cases which are assigned as positive (1, here) and is actually observed to be 0 to the total number of actual negatives. AUROC can take values between 0 and 1 and larger values indicate better performance. \\

\subsubsection{Continuous data accuracy measures}

\noindent We preferred one scale dependent and one scale independent accuracy measures for continuous data, following Hyndman (2006) and Hyndman and Koehler (2006). The scale dependent measure is mean absolute error (MAE) which can be calculated by

\begin{align}
\mbox{MAE}_t = \frac{1}{N} \sum_{i=1}^{N} |e_{it}|,
\end{align}

\noindent where $e_{it}$ is the forecast error and calculated by $e_{it}=X_{it}-\hat{X}_{it}$. MAE has some advantages over the other scale dependent measures, e.g. mean squared (MSE) and root mean squared (RMSE) errors. For instance, MAE is in the same scale of data (unlike MSE) and is not heavily affected from outliers (unlike MSE and RMSE; Hyndman and Koehler, 2006).

The scale independent accuracy measure is mean absolute scaled error (MASE) proposed by Hyndman and Koehler (2006). Related calculation is given by

\begin{align}
\label{eq:mase2}
\mbox{MASE}_t=\frac{1}{N} \sum_{i=1}^{N} \frac{|e_{it}|} {\frac {1} {T-1} \sum_{h=2}^{T} |X_{ih}-X_{i,h-1}| } .
\end{align}

Lower values of MAE and MASE indicate better model performance. \\

\section{Forecasting mother's stress and children's morbidity}

\subsection{Data}

\noindent In MSCM study, 167 mothers and their pre-school children (aged between 18 months-5 years) were enrolled mainly to understand the relationship between mother's employment status and the pediatric care usage. In a baseline interview, some demographic and family information were collected through the following variables: the marriage status, education level and employment status of mothers, the health status of mothers and children at baseline and the size of the household (Table 1). After the baseline interview, mothers were asked to keep the records of their stress and their children's illness statuses with a 28-day health diary. These variables were dichotomised later as stress status of mothers (stress: 0=absence, 1=presence) and illness status of children (illness: 0=absence, 1=presence).\\

\vspace{0.05in} 
\noindent \textcolor{red}{Table 1 is about here.}  \\
\vspace{0.05in}

Empirical investigations of the within-subject association structures of both responses suggested extremely weak within-response dependencies in the period of days 1 to 16. Therefore, we only considered the period of days 17 to 28. Nonetheless, we calculated the averages of responses for the period of days 1 to 16, and considered these as two new independent variables (bstress and billness, respectively in Table 1), to capture the individual characteristics of mothers and their children. Standardised time (week in Table 1) and the interactions between time and some independent variables were included as additional covariates.

We partitioned the MSCM data set (days 17 to 28) into two: 1) model building (days 17 to 24), and 2) forecast validation (days 25 to 28) parts. Once the aforementioned models were built for the former time period, forecastings were done based on the systems indicated by these models.\\ 

\subsection{Results}

\noindent We checked the existence of multicollinearity problems by variance inflation factor, via pooled (over time points) logistic regression models. Results (not shown here) showed that none of the related values were greater than 1.394 which indicated no multicollinearity problems. 

We built several models to explain the MSCM data set in the period of days 17 to 24. Due to page limits, we could not include the modelling results here, but provide some details below. There is poverty in model selection with GEE, since it does not define a genuine likelihood function. Therefore, we built several UMM's and MMM1's with different independent variable sets and working variance-covariance structure assumptions. We did forecasting with each of these models and best results were reported here. On the other hand, we built a model with MMM2 based on the results of MMM1. This MMM2 permitted estimating five less parameters compared to its mother MMM1. We were able to use well established model selection methods, e.g. likelihood ratio test for MMREM and PNMTREM(1), since the parameters of these models were estimated via MLE. Forecastings were done by these best models. 

We reported some features of the models in Table 2, including the names and availability of the related R (R Core Development Team, 2013) packages and computational details. The model building processes of UMM, MMM1 and MMM2 took very short computing times. On the other hand, MMREM and PNMTREM(1) took more computational times for the same data set (MSCM). All of these models were fitted on a personal computer with 4.00 GB RAM and 2.53 GHz processor.  \\

\vspace{0.05in} 
\noindent \textcolor{red}{Table 2 is about here.}  \\
\vspace{0.05in}

As it was mentioned earlier, we considered generating $z_i$'s from independent standard normal distributions $K$ times while forecasting with MMREM and call this method as MMREM3. We considered $K=30, 50, 80,$ $100, 120, 150, 180, 200, 250, 300, 400, 500, 750, 1000$ and calculated percentage differences in accuracy measures for successive replication amounts, e.g. $K=30$ vs $K=50$. Related percentage differences (not shown here) directed us to choose $K=150$ for response=stress and $K=250$ for response=illness. The reason for greater $K$ for response=illness might be due to the fact that the response has a greater variance. For instance, the variance parameter estimates of stress and illness were found to be 2.07 and 4.56 by MMREM, respectively.

We observed that the choice of how to handle $Y_{i,t-1,j}$ in \eqref{eq:pnmtremlev2} at $day=26, 27$ and $28$ while forecasting with PNMTREM(1) is decisive rather than the choice of using $\hat{z}_i$ or assigning $z_i$. As expected using the observed responses in place of dichotomizing $Y_{i,t-1,j}$ yielded the best results. However, we preferred not to report these results here since this approach does not reflect the real life cases in which we do not observe these responses for these time points. The second best results were attained by the use of $c_j=0.5$ to dichotomise $Y_{i,t-1,j}$ and the results of this approach were reported here, i.e., the results of PNMTREM1 and PNMTREM2. Simulating the responses from Bernoulli distributions followed this approach. Empirical $c_j$ yielded very poor results, especially in terms of ePCP.

Based on several accuracy measures, e.g. RMSE and MAE, and several model selection criterion, e.g. Akaike Information Criterion and Bayesian Information Criterion, we used exponential smoothing with additive error, no trend, no seasonality to forecast $\alpha_{t}$ and exponential smoothing with additive error, additive trend, no seasonality to forecast $\sigma_t$. 

Model building (days 17 to 24) and forecasting (days 25 to 28) results of mother's stress and children's illness are presented in Tables 3 and 4, respectively. For model building period, the marginalised models, namely, MMREM and PMTREM(1) seemed to perform better compared to UMM, MMM1 and MMM2. For instance, for response=stress MMREM3 and MMREM4 outperformed these marginal models; the corresponding ePCP values were 0.844 and 0.842 for MMREM's versus 0.799 and 0.800 for the marginal models. Moreover, the AUROC values of MMREM1 and MMREM2 were found to be 0.82 as opposed to 0.72 for marginal models. Note that MMREM1 and MMREM2 were identical models at days 17 to 24 indeed. PNMTREM1 followed these models with an AUROC value of 0.804. Similar model ranking was observed for response=illness. For instance, in terms of ePCP all the MMREM's and PNMTREM1 seemed to outperform the others. In terms of AUROC, MMREM1, MMREM2 and PNMTREM1 seemed to be the best models.

For forecasting time period, marginally specified models also outperformed the marginal models for both of the responses. In terms of ePCP, MMREM's seemed to be the best performing models. For instance, the ePCP values of MMREM4 were all higher than 0.860 and for response=illness at day 28 it was found to be 0.917. PNMTREM's seemed to be the worst performing ones in terms of this accuracy measure for the forecasting time periods. This worst performance of PNMTREM's is less apparent for response=stress, which is the response with lower variance. In terms of AUROC, PNMTREM's, especially PNMTREM2, outperformed the other models. For instance, the AUROC value of this method for response=stress at day 27 was found to be 0.843. On the other hand, the AUROC values for other methods were lower; the lowest was found to be 0.719 belonging to MMREM4. UMM and MMM's were among the worst performing methods in terms of both ePCP and AUROC in most cases and these models had similar forecasting performances. \\

\vspace{0.05in} 
\noindent \textcolor{red}{Table 3 is about here.}  
\vspace{0.05in}

\vspace{0.05in} 
\noindent \textcolor{red}{Table 4 is about here.}  
\vspace{0.05in} \\

\section{A simulation assessment}

\noindent Forecasting results of MSCM data set suggested that marginalised models yielded improved forecasts. Additionally, the results of UMM seemed to be worst in general, yet these results were close to the ones for MMM's. This results was our expectation. However empirical investigations of dependence between responses suggested a weak dependence between stress and illness at the period of days 17 to 28, e.g. Spearman rank correlation was found to be 0.13. Based on these, we needed to investigate the forecasting performances of the models under different scenarios and conducted a simulation study.\\

\subsection{Data generation}

\noindent In the simulation study, we assumed that there were 500 subjects $(i=1, \ldots, 500)$ who were followed repeatedly over 8 time points $(t=1, \ldots, 8)$. At each follow up, six different variables were assumed to be collected. Among them, two were assumed to be the responses $(k=1,2)$ and four were assumed to be the covariates. Among the covariates, while two ($X_1$ and $X_3$) were assumed to be time invariant, other two ($X_2$ and $X_4$) were assumed to be time varying. We mainly considered that all of the variables, including the responses, were continuous and generated them from a multivariate normal distribution with a specified mean and variance-covariance combination. Specifically, all the variables were assumed to have mean 0. Moreover, we assumed that while the continuous versions of the response variables, $Y_1^*$ and $Y_2^*$, had variances of 1.5 and 2.5, the explanatory variables, $X_1, X_2, X_3$ and $X_4$, had variances of 8, 2.5, 15 and 25, respectively. We further assumed high autocorrelations for responses; as well as high correlations between responses and covariates; mild correlations between responses; and low correlations between covariates (Table 5). Although the correlations in Table 5 seemed quite high, e.g. $cor(Y^*_t, Y^*_{t-1})=0.9$, they decreased after data generation and dichotomization to obtain the response variables. The continuous versions of the response variables were dichotomised according to the following rule: classify $Y_{itj}$ as 1 if $Y^*_{itj} \geq 0$, and 0 otherwise.\\

\vspace{0.05in} 
\noindent \textcolor{red}{Table 5 is about here.}  
\vspace{0.05in} \\

\subsection{Results}

\subsubsection{Independent variables}

\noindent As it was stated earlier, we first forecast the independent variables. Results for $X_2$ are presented in Table 6. These results were calculated over 10,000 replications and obtaining all of the them took 26.3 minutes. Mean and standard error (SE) of 10,000 MAE and MASE values were calculated and reported in this table.\\

\vspace{0.05in} 
\noindent \textcolor{red}{Table 6 is about here.}  \\
\vspace{0.05in}

MAE values indicated that TM(1) and TM(2) performed very similar in terms of forecasting $X_2$. While the mean values were very close to each other, standard errors were same for forecasting period. For instance, whereas for $t=6$ the mean MAE was found to be 0.917 for TM(1), it was found to be 0.914 for TM(2). Moreover, their standard errors were found to be 0.031. Note that although the standard errors of MAE of TM(1) and TM(2) for model building periods seemed to be slightly different, they were not directly comparable, since while the former considers $2$nd, $3$rd and $4$th time points in the model building period, i.e., 1,500 observations, the latter considers only $3$rd and $4$th time points, i.e., 1,000 observations. MASE values indicated that TM(1) performed better in terms of forecasting $X_2$ compared to TM(2) (Table 6). 

Recall that while $X_2$ was assumed to have a variance of 2.5 during the simulation process, $X_4$ had a variance of 25. However, increase in the variance of the independent variables did not seem to effect the performances of TM(1) and TM(2). Similar to $X_2$, while MASE results of $X_4$ indicated that TM(1) performed better in terms of forecasting $X_4$ compared to TM(2), MAE results indicated similar performance (results are not shown here).

Notice that we estimated one more parameter in TM(2) compared to TM(1) and the estimation of this extra parameter seemed to be redundant since it did not contribute to the forecasting accuracy. Therefore, TM(1) was preferred in terms of forecasting the independent variables during our simulation studies. These results were in agreement with the ones reported in Aslan (2010) in which the results were based only on mean squared error (MSE).\\

\subsubsection{Dependent variables}

\noindent After forecasting the independent variables, we considered forecasting the bivariate longitudinal binary responses, $Y_1$ and $Y_2$. Based on the forecasting performances of the models on MSCM data set, we selected UMM, MMM1, MMREM2, MMREM4, PNMTREM1 and PNMTREM2. We did not consider MMM2 for the simulation study on forecasting, since the results of MMM1 and MMM2 seemed to be very similar for the MSCM data set. Moreover, model fitting of MMM2 requires more computational time compared to MMM1 (Table 2).

Simulation results on forecasting $Y_1$ are presented in Table 7. We preferred not to show the results of $Y_2$ here due to page limits. Indeed, these results seemed to be slightly better than the ones for $Y_1$, but indicated same model rankings. The simulation study were replicated 100 times and one replication (the last one) took 11.7 minutes. Mean and standard error (SE) of the ePCP and AUROC values of these 100 simulation replications were reported.\\

\vspace{0.05in} 
\noindent \textcolor{red}{Table 7 is about here.}  \\
\vspace{0.05in}

In the model building period, times 1 to 4, MMREM2 and PNMTREM1 seemed to be the best methods in terms both ePCP and AUROC. For instance, for $Y_1$ while the ePCP and AUROC values of the former model were found to be 0.811 and 0.924, these values for the latter model were found to be 0.766 and 0.905 (Table 7). These results seemed to be even better for $Y_2$, especially for MMREM2; the corresponding ePCP and AUROC values for MMREM2 were found to be 0.872 and 0.966. For forecasting time periods, ePCP and AUROC indicated different best models similar to the forecasting with MSCM data set. Specifically, ePCP indicated that MMREM2 \& MMREM4 were the best performed methods. For instance, for $Y_1$ at time 6, while the ePCP values of MMREM2 \& MMREM4 were found to be 0.708 and 0.670, respectively, these values were found to be 0.605 for both UMM \& MMM1, and 0.566 and 0.555 for PNMTREM1 \& PNMTREM2. In terms of AUROC, MMREM2 and PNMTREM1 seemed to be the best models. For instance, for $Y_1$ at time 8, the AUROC values of these models were found to be 0.734 and 0.755, respectively. On the other hand, while the AUROC values of UMM, MMM1 \& MMREM4 were found to be 0.683, it was found to be 0.706 for PNMTREM2. The standard errors of ePCP and AUROC values for all of the models seemed to be similar. 

For one-step forecasts, i.e., time=5, PNMTREM's seemed to perform similar to UMM and MMM1 in terms of ePCP. Moreover, at this time point, PNMTREM's seemed to be better than MMREM's in terms of AUROC. For instance, for $Y_1$, the AUROC values of PNTREM1 and MMREM2 were 0.891 and 0.812, respectively.  

To sum up, complex models, especially MMREM2 and PNMTREM1, outperformed in model building process. It was observed that MMREM2 was one of the leading methods in both modelling and forecasting procedure, and was computationally efficient as well. However, PNMTREM's, especially PNMTREM1, seemed to be promising in terms of one-step ahead forecasts.\\

\section{Conclusion and discussion}

\noindent In this study, we considered forecasting multivariate longitudinal binary data. We mainly considered the comparison of five different models in terms of forecasting such data. Among them whereas two (MMREM and PNMTREM(1)) are complex models, the others (UMM, MMM1 and MMM2) are relatively simpler ones. The forecasting abilities of these models were assessed via a real life data (MSCM) and a simulation study. The simulation datasets were generated under a model independent scenario to provide a fair model competition. To best mimic the real life forecasting problems, we considered forecasting the independent variables as well as the dependent ones. The forecasts, for both dependent and independent variables, were checked via several accuracy measures.

Both real life and simulation examples showed that complex models outperformed during the model building process in terms of all the accuracy measures. For the forecasting period, the complex models again outperformed the simpler models. We observed that while AUROC suggests PNMTREM(1), ePCP suggests MMREM as the best forecasting method. This is most probably due to different characteristics of these accuracy measures. In other words, these measures consider different aspects of forecasts. To illustrate, while ePCP considers the averaged probability with respect to actual observations, AUROC considers classification of the forecast probabilities according to several threshold values. Based on these, it can be said that the latter is more robust to threshold values. Hence, we can conclude that while the forecast results of PNMTREM(1) are more robust, MMREM results are better on the average. UMM and MMM's were among the worst performing methods in terms of both ePCP and AUROC. Surprisingly, UMM and MMM's performed similar in terms of both model building and forecasting. Moreover, these complex models are now available to the practitioners via R packages or codes.

The computational efficiencies of models in terms of parameter estimation and forecasting procedures are also important in real life. We observed that while the marginal models took a few seconds for parameter estimation, the marginalised models took longer computational times. Moreover, the forecasting procedures of the former models are quite simpler compared to the latter ones. For instance, we do not need to forecast extra time varying parameters for the former ones, i.e., they avoid the use of numerical methods and exponential smoothing. Nevertheless, the gains in the forecasts directed us to prefer the complex models for forecasting purposes. 

The use of the forecast values of the independent variables, instead of using supposedly the observed ones, yielded decreases in the accuracy measures. However, this did not change the model rankings (Aslan, 2010). We observed that the response variance was affecting the forecasting results, but it did not change the model rankings. Specifically, we observed higher accuracy measures for the responses with higher variances, i.e., children's illness in the MSCM data set and $Y_2$ in the simulation study. Nevertheless, they indicated same model rankings with the mother's stress and $Y_1$. 

In this study, we mainly considered forecasting bivariate response. A natural extension of it might be forecasting more than two responses.

\newpage


\begin{table}
\centering
\caption{List of the variables in the MSCM study and the related explanations.}
\label{tab:mscmvar}
\fbox{
\begin{tabular} {l l}
Variable & Explanation\\
\hline
stress & mother's stress status: 0=absence, 1=presence \\
illness & child's illness status: 0=absence, 1=presence \\
married & marriage status of the mother: 0=other, 1=married \\
education & mother's education level: 0=less than high school, \\
 & 1=at least high school graduate \\
employed & mother's employment status: 0=unemployed, 1=employed \\
chlth & child's health status at baseline: 0=very poor/poor, \\
 & 1=fair, 2=good, 3=very good \\
mhlth & mother's health status at baseline: 0=very poor/poor, \\
 & 1=fair, 2=good, 3=very good \\
housize & size of the household: 0=2-3 people, 1=more than 3 people\\
bstress & baseline stress: average value of the mother's stress \\
 & status for the first 16 days \\
billness & baseline illness: average value of the child's illness \\
 & status for the first 16 days \\
week & a time variable calculated as (day-22)/7\\
mhlth*week & interaction between mother's health status at baseline and time \\
housize*week & interaction between size of the household and time \\
billness*week & interaction between baseline illness and time \\
\end{tabular}}
\end{table}  

\newpage

\begin{table}
\centering
\caption{Model building with MSCM data set.}
\label{tab:mscm.mod.build.feat}
\fbox{
\scalebox{0.85}{
\begin{tabular}{l l l l l}
Model       &  Package     & Function                 &  Reference           & Computational Time for MSCM  \\ \hline
UMM         &  gee         &  gee                     & Carey (2012)       & $<$ 1 sec \\
MMM1        &  mmm         &  mmm                     & Asar and Ilk (2013)     & $<$ 1 sec \\
MMM2        &  mmm2        &  mmm2                    & Asar and Ilk (2014)     & $<$ 1 sec \\ 
MMREM       &  From author &  findmle + FORTRAN dll's & Lee et al. (2009)      & $\approx$ 7-8 mins \\ 
PNMTREM(1)  &  pnmtrem     &  pnmtrem1                & Asar et al. (2014)    & $\approx$ 30-40 mins \\ 
\end{tabular}}}
\end{table}  

\newpage

\begin{table}
\centering
\caption{Forecast results for mothers' stress.}
\label{tab:stress.forecast}
\fbox{
\begin{tabular}{l c c c c c c}
Model &  Day   & ePCP  &  AUROC  & Day   & ePCP  &  AUROC \\ \hline
UMM \ \ \ (Exch)   	      & \multirow{9}{*}{17 to 24} & 0.799 & 0.726 & \multirow{9}{*}{27} & 0.858 & 0.790  \\
MMM1 (Exch)     	      &   						  & 0.799 & 0.726 &					  &	0.855 & 0.747  \\
MMM2 (Exch)      	      &    						  & 0.800 & 0.722 &				   	  & 0.861 & 0.780  \\ 
MMREM1 \ \ \ \ \ \ \  	  &   						  & 0.828 & 0.821 &					  & 0.851 & 0.757  \\
MMREM2  \ \ \ \ \ \ \  	  &   						  & 0.828 & 0.821 &					  & 0.884 & 0.754  \\	
MMREM3 \ \ \ \ \ \ \  	  &    						  & 0.844 & 0.712 &					  & 0.903 & 0.735  \\ 
MMREM4 \ \ \ \ \ \ \  	  &  						  & 0.842 & 0.721 &				   	  & 0.900 & 0.719  \\
PNMTREM1 \ \ \ \      	  & 						  & 0.824 & 0.804 &				   	  & 0.662 & 0.764    \\
PNMTREM2 \ \ \ \      	  &   						  & 0.829 & 0.712 &				  	  & 0.754 & 0.843   \\ \hline
UMM \ \ \ (Exch)	   	  & \multirow{9}{*}{25}       & 0.823 & 0.678 & \multirow{9}{*}{28} & 0.846 & 0.709   \\
MMM1 (Exch)      	 	  & 						  & 0.821 & 0.641 &				  	  & 0.842 & 0.670   \\
MMM2 (Exch)               & 						  & 0.823 & 0.657 &				  	  & 0.845 & 0.690   \\
MMREM1 \ \ \ \ \ \ \      & 						  & 0.831 & 0.775 &				  	  & 0.830 & 0.585   \\
MMREM2 \ \ \ \ \ \ \      & 						  & 0.866 & 0.759 &				  	  & 0.859 & 0.595   \\
MMREM3 \ \ \ \ \ \ \      & 						  & 0.868 & 0.675 &				  	  & 0.899 & 0.661   \\ 
MMREM4 \ \ \ \ \ \ \      & 						  & 0.867 & 0.695 &				  	  & 0.886 & 0.651   \\
PNMTREM1 \ \ \ \          & 						  & 0.688 & 0.761 &				  	  & 0.625 & 0.608    \\
PNMTREM2 \ \ \ \          & 						  & 0.723 & 0.743 &				  	  & 0.732 & 0.735   \\ \hline
UMM \ \ \ (Exch)          & \multirow{9}{*}{26}       & 0.829 & 0.704 & \multirow{9}{*}{25 to 28} & 0.839 & 0.718  \\
MMM1 (Exch)               & 						  & 0.828 & 0.704 &				  	  & 0.841 & 0.687  \\
MMM2 (Exch)               & 						  & 0.829 & 0.715 &				  	  & 0.840 & 0.708  \\
MMREM1 \ \ \ \ \ \ \      & 						  & 0.836 & 0.811 &				  	  & 0.837 & 0.729  \\
MMREM2 \ \ \ \ \ \ \      & 						  & 0.865 & 0.803 &				  	  & 0.868 & 0.711  \\
MMREM3 \ \ \ \ \ \ \      & 						  & 0.871 & 0.728 &				  	  & 0.883 & 0.699  \\ 
MMREM4 \ \ \ \ \ \ \      & 						  & 0.869 & 0.725 &				  	  & 0.880 & 0.698  \\
PNMTREM1 \ \ \ \          & 						  & 0.685 & 0.819 &				  	  & 0.665 & 0.736  \\
PNMTREM2 \ \ \ \          & 						  & 0.711 & 0.731 &				  	  & 0.730 & 0.759  \\ 
\end{tabular}} \\
Note: MMREM1 and MMREM2 are identical for model building periods.
\end{table}

\newpage

\begin{table}
\centering
\caption{Forecast results for children's illness.}
\label{tab:illness.forecast}
\fbox{
\begin{tabular}{l c c c c c c }
Model &  Day   & ePCP  &  AUROC  &  Day   & ePCP  &  AUROC    \\ \hline
UMM \ \ (AR-1)    	 & \multirow{9}{*}{17 to 24} & 0.815 & 0.719 &  \multirow{9}{*}{27}  &  0.793  & 0.560      \\
MMM1 (Exch)    	     &                           & 0.815 & 0.720 &  			     	 &  0.800  & 0.562      \\
MMM2 (Exch)          &                           & 0.815 & 0.719 & 					     &  0.800  & 0.575      \\
MMREM1 \ \ \ \ \ \ \ &                           & 0.868 & 0.878 &  					 &  0.759  & 0.656      \\
MMREM2 \ \ \ \ \ \ \ &                           & 0.868 & 0.878 & 					     &  0.873  & 0.677      \\
MMREM3 \ \ \ \ \ \ \ &                           & 0.881 & 0.697 &  					 &	0.907  & 0.662      \\ 
MMREM4 \ \ \ \ \ \ \ &                           & 0.882 & 0.705 &  					 &	0.908  & 0.665      \\
PNMTREM1 \ \ \ \     &                           & 0.855 & 0.814 &  					 &	0.526  & 0.686      \\
PNMTREM2 \ \ \ \     &                           & 0.851 & 0.691 &  					 &	0.605  & 0.741      \\ \hline
UMM \ \ (AR-1)       & \multirow{9}{*}{25}       & 0.779 & 0.640 &  \multirow{8}{*}{28}  &	0.799  & 0.687      \\
MMM1 (Exch)          &                           & 0.782 & 0.632 &  					 &	0.807  & 0.685      \\
MMM2 (Exch)          &                           & 0.781 & 0.608 &  					 &	0.805  & 0.674      \\
MMREM1 \ \ \ \ \ \ \ &                           & 0.785 & 0.736 &  					 &	0.761  & 0.622      \\
MMREM2 \ \ \ \ \ \ \ &                           & 0.856 & 0.674 &  					 &	0.853  & 0.647      \\
MMREM3 \ \ \ \ \ \ \ &                           & 0.857 & 0.570 &   					 &	0.911  & 0.683      \\ 
MMREM4 \ \ \ \ \ \ \ &                           & 0.860 & 0.582 &  					 &	0.917  & 0.686      \\
PNMTREM1 \ \ \ \     &                           & 0.629 & 0.798 &  					 &	0.514  & 0.638      \\
PNMTREM2 \ \ \ \     &                           & 0.652 & 0.783 &  					 &	0.605  & 0.772      \\ \hline 
UMM \ \ (AR-1)       & \multirow{9}{*}{26}       & 0.797 & 0.623 &  \multirow{9}{*}{25 to 28} & 0.792  & 0.617  \\
MMM1 (Exch)     	 &                           & 0.802 & 0.627 &  					 &	0.798  & 0.616      \\
MMM2 (Exch)    	     &                           & 0.801 & 0.614 &  					 &	0.800  & 0.608      \\
MMREM1 \ \ \ \ \ \ \ &                           & 0.764 & 0.677 &  					 &	0.767  & 0.675	    \\
MMREM2 \ \ \ \ \ \ \ &                           & 0.875 & 0.725 &  					 &	0.864  & 0.657      \\
MMREM3 \ \ \ \ \ \ \ &                           & 0.893 & 0.670 &  					 &	0.892  & 0.636      \\ 
MMREM4 \ \ \ \ \ \ \ &                           & 0.895 & 0.688 &  					 &	0.895  & 0.645      \\
PNMTREM1 \ \ \ \     &                           & 0.555 & 0.713 &  					 &	0.556  & 0.701      \\
PNMTREM2 \ \ \ \     &                           & 0.624 & 0.791 &  					 &	0.621  & 0.765 	    \\ 
\end{tabular}} \\
Note: MMREM1 and MMREM2 are identical for model building periods.
\end{table}

\newpage

\begin{table}
\centering
\caption{A summary of the assumed correlation structure.}
\label{tab:sim.study.corstr}
\fbox{
\begin{tabular} {c c c c c c}
Time lag  &  $Y^*_t, Y^*_{t^{\prime}}$ & $X_t, X_{t^{\prime}}$  & $Y^*_t, X_t$ & $Y^*_{t1}$, $Y^*_{t2}$ & $X_{tl}$, $X_{tl^{\prime}}$  \\ \hline 
0 & 1.00 & 1.00 & 0.80 & 0.60 & 0.20 \\
1 & 0.90 & 0.88 & 0.70 & 0.55 & 0.18 \\
2 & 0.80 & 0.76 & 0.60 & 0.45 & 0.16 \\
3 & 0.70 & 0.64 & 0.50 & 0.40 & 0.14 \\
4 & 0.60 & 0.52 & 0.40 & 0.35 & 0.12 \\
5 & 0.50 & 0.40 & 0.30 & 0.30 & 0.10 \\
6 & 0.40 & 0.28 & 0.20 & 0.25 & 0.08 \\
7 & 0.30 & 0.16 & 0.10 & 0.20 & 0.06 \\
\end{tabular}}
\end{table}

\newpage

\begin{table}
\centering
\caption{Forecasting results of $X_2$.}
\label{tab:x2.for.results}
\fbox{
\begin{tabular} {c c c c c c c c c c}
\multicolumn{1}{c}{} & \multicolumn{4}{c}{TM(1)} & \multicolumn{1}{c}{} & \multicolumn{4}{c}{TM(2)}\\ \hline 
\multicolumn{1}{c}{} & \multicolumn{2}{c}{MAE} & \multicolumn{2}{c}{MASE} & \multicolumn{1}{c}{} & \multicolumn{2}{c}{MAE} & \multicolumn{2}{c}{MASE} \\ \cline{2-5} \cline{7-10} 
Time  & Mean  & SE    & Mean  & SE    & Time & Mean     & SE  & Mean     & SE        \\ \hline 
2 to 4   & 0.671 & 0.013 & 0.993 & 0.012 & 3 to 4  & 0.668 & 0.016 & 1.046 & 0.057 \\
5     & 0.671 & 0.023 & 1.242 & 0.068 & 5    & 0.670 & 0.023 & 1.531 & 0.175 \\
6     & 0.917 & 0.031 & 1.692 & 0.092 & 6    & 0.914 & 0.031 & 2.086 & 0.326 \\
7     & 1.084 & 0.037 & 1.996 & 0.111 & 7    & 1.079 & 0.037 & 2.457 & 0.317 \\
8     & 1.208 & 0.041 &  2.223 &  0.123 & 8    & 1.201 & 0.041 &  2.728 &  0.335 \\
5 to 8   & 0.970 & 0.025 & 1.788 & 0.085 & 5 to 8  & 0.966 & 0.025 & 2.200 & 0.262 \\
\end{tabular}} \\
Note: $X_2$ had variance of 2.5. Results were calculated over 10,000 replications.
\end{table}

\newpage

\begin{table}
\centering
\caption{Forecasting results of $Y_1$ over 100 replications.}
\label{tab:y1.for.results}
\scalebox{0.85}{
\fbox{
\begin{tabular} {l c c c c c c c c c c}
\multicolumn{1}{c}{} & \multicolumn{1}{c}{} & \multicolumn{2}{c}{ePCP} & \multicolumn{2}{c}{AUROC} & \multicolumn{1}{c}{}  & \multicolumn{2}{c}{ePCP} & \multicolumn{2}{c}{AUROC} \\ \cline{3-6} \cline{8-11}
Model   				& Time                    & Mean  & SE    & Mean  & SE    & Time                  & Mean  & SE    & Mean  & SE     \\ \hline 
UMM \ \ \ (Uns)         & \multirow{6}{*}{1 to 4} & 0.648 & 0.011 & 0.810 & 0.013 & \multirow{6}{*}{7}    & 0.585 & 0.012 & 0.713 & 0.025  \\
MMM1 \ (Uns)            &                         & 0.649 & 0.011 & 0.810 & 0.014 &						  & 0.585 & 0.012 & 0.713 & 0.025  \\
MMREM2 \ \ \ \ \        & 						  & 0.811 & 0.011 & 0.924 & 0.008 &						  & 0.683 & 0.018 & 0.768 & 0.023  \\
MMREM4 \ \ \ \ \ 		& 					   	  & 0.721 & 0.013 & 0.810 & 0.014 &						  & 0.644 & 0.019 & 0.712 & 0.024  \\
PNMTREM1 \              & 						  & 0.766 & 0.009 & 0.905 & 0.008 &  				      & 0.552 & 0.021 & 0.764 & 0.023 \\
PNMTREM2 \              &                         & 0.740 & 0.010 & 0.870 & 0.009 & 					  & 0.544 & 0.020 & 0.714 & 0.025  \\ \hline
UMM \ \ \ (Uns)         & \multirow{6}{*}{5}      & 0.625 & 0.012 & 0.778 & 0.019 &  \multirow{6}{*}{8}   & 0.569 & 0.012 & 0.683 & 0.027 \\
MMM1 \ (Uns)            & 						  & 0.626 & 0.012 & 0.778 & 0.019 &						  & 0.569 & 0.012 & 0.683 & 0.027  \\
MMREM2 \ \ \ \ \ 		& 						  & 0.720 & 0.016 & 0.812 & 0.017 &						  & 0.658 & 0.019 & 0.734 & 0.026  \\
MMREM4 \ \ \ \ \		& 						  & 0.694 & 0.017 & 0.778 & 0.019 & 				      & 0.622 & 0.020 & 0.683 & 0.027 \\
PNMTREM1 \       		& 						  & 0.639 & 0.016 & 0.891 & 0.015 &					   	  & 0.547 & 0.021 & 0.755 & 0.029  \\
PNMTREM2 \       		& 						  & 0.622 & 0.016 & 0.884 & 0.016 & 					  & 0.539 & 0.020 & 0.706 & 0.029 \\ \hline
UMM \ \ \ (Uns)         & \multirow{6}{*}{6}      & 0.605 & 0.011 & 0.747 & 0.020 & \multirow{6}{*}{5 to 8} & 0.596 & 0.011 & 0.732 & 0.019 \\
MMM1 \ (Uns)            & 						  & 0.605 & 0.011 & 0.747 & 0.020 & 					  & 0.596 & 0.010 & 0.732 & 0.019 \\
MMREM2 \ \ \ \ \ 		& 				  		  & 0.708 & 0.016 & 0.800 & 0.018 & 					  & 0.692 & 0.013 & 0.779 & 0.016 \\
MMREM4 \ \ \ \ \ 		& 						  & 0.670 & 0.016 & 0.747 & 0.020 & 					  & 0.658 & 0.015 & 0.732 & 0.018 \\
PNMTREM1 \       		& 						  & 0.566 & 0.020 & 0.802 & 0.019 & 					  & 0.573 & 0.016 & 0.801 & 0.016 \\
PNMTREM2 \       		& 						  & 0.555 & 0.020 & 0.748 & 0.021 & 					  & 0.563 & 0.016 & 0.761 & 0.017 \\				
\end{tabular}}}	\\
Note: Uns denotes unstructured working variance-covariance structure assumption.
\end{table}

\end{document}